\documentclass[twocolumn,aps,prl,superscriptaddress]{revtex4}
\usepackage{color}
\usepackage{mathtools}
\usepackage{graphicx}
\usepackage{SIunits}

\begin{document}

\title{{Tunable negative permeability in a three-dimensional superconducting metamaterial}}
\author{C. Kurter}
\affiliation{Center for Nanophysics and Advanced Materials, Department of Physics, University of Maryland, College Park, MD 20742-4111}
\affiliation{Department of Physics, Missouri University of Science and Technology, Rolla, MO 65409}

\author{T. Lan}
\affiliation{Center for Nanophysics and Advanced Materials, Department of Physics, University of Maryland, College Park, MD 20742-4111}
\affiliation{Department of Materials Science and Engineering, Northwestern University, Evanston, IL 60208, USA}

\author{L. Sarytchev}

\affiliation{Center for Nanophysics and Advanced Materials, Department of Physics, University of Maryland, College Park, MD 20742-4111}

\author{Steven M. Anlage}
\affiliation{Center for Nanophysics and Advanced Materials, Department of Physics, University of Maryland, College Park, MD 20742-4111}
\affiliation{Department of Electrical and Computer Engineering, University of Maryland, College Park, MD 20742-3285}

\date{\today}

\begin{abstract}

We report on highly tunable radio frequency (rf) characteristics of a low-loss and compact three dimensional (3D) metamaterial made of superconducting thin film spiral resonators. The rf transmission spectrum of a single element of the metamaterial shows a fundamental resonance peak at $\sim$24.95 MHz that shifts to a 25$\%$ smaller frequency and becomes degenerate when a 3D array of such elements is created. The metamaterial shows an $\emph{in-situ}$ tunable narrow frequency band in which the real part of the effective permeability is negative over a wide range of temperature, which reverts to gradually near-zero and positive values as the superconducting critical temperature is approached. This metamaterial can be used for increasing power transfer efficiency and tunability of electrically small rf-antennas. 

\end{abstract}

\maketitle

\section{Introduction}

A three dimensional (3D) array of sub-wavelength elements composing a metamaterial can demonstrate unique electromagnetic properties which are not directly accessible in nature, such as negative permeability~\cite{Pendry99, PhysRevLett.84.4184} and zero refractive index~\cite{Silveirinha_zero_index, Ziolkowski_zero_index}. Most metamaterials achieve these unusual effects by means of tailored geometrical structures without relying on the intrinsic properties of materials used in metamaterial fabrication. Conducting non-magnetic split-ring resonators (SRRs), and their derivatives, have been employed to demonstrate negative permeability by strongly coupling to the magnetic field component of incident electromagnetic waves at microwave~\cite{Shelby} or higher frequency bands~\cite{Hussain, Yen05032004}. However, when lower operational frequencies are needed, these resonator-based metamaterials require impractically large dimensions to be able to compensate for ohmic losses~\cite{KurterAPL2010}. One way to address this miniaturization problem is to use superconductors since they show very low surface resistance below their transition temperature~\cite{Anlagereview2011, Anlagereview2014}. Moreover, tunability of the resonant features can be precisely achieved without lumped elements by means of temperature and magnetic field~\cite{RicciIEEE, KurterAPL2012, PhysRevX.3.041029}.  For instance, the plasma frequency of superconducting spiral meta-atoms have been tuned to the MHz range by means of temperature variation, and the plasmonic properties have been studied at essentially fixed frequency~\cite{PhysRevB.88.180510}.

Of particular interest are radio frequency (rf) metamaterials where the free space wavelength $\lambda_0$ can be tens of meters. Despite the challenge of scaling the dimensions down, rf metamaterials offer a wide spectrum of applications including high resolution magnetic resonance imaging~\cite{Wiltshire, Freire2008, Freire2010}, magneto-inductive lenses~\cite{Freire} and wireless communication ~\cite{Kurs, Urzhumov, Sedwick, Wang} employing electrically small antennas. Antennas integrated with compact rf metamaterials were speculated to have enhanced radiation power, high directivity and reduced return loss due to the unique electromagnetic properties of metamaterials~\cite{BilottiAntenna, Werner}. Ziolkowski~\cite{Ziolkowski2003, Ziolkowski2006} has proposed using a metamaterial with negative refractive index to boost the real radiated power of an electrically short dipole antenna (size $\lambda_0$/1000) by almost two orders of magnitude compared to the dipole-in-free-space case. A large imaginary part (reactance) of radiation impedance Z$_{rad}$=R$_{rad}$+iX$_{rad}$ of an electrically small antenna limits the radiation efficiency because free space has a purely real impedance~\cite{Hansen}. When the antenna is placed in near-field proximity to a negative refractive index medium, the X$_{rad}$ of the antenna can be cancelled out. Therefore the metamaterial acts as a natural impedance matching network for the dipole, which maximizes the transfer of power from the antenna to free space in the far-field. The bandwidth and the operating frequency of the antennas can also be reconfigured with tunable metamaterials, which is crucial for modern wireless communication systems relying on multi-band operation~\cite{Werner}. 

In this work, we demonstrate a tunable 3D rf metamaterial composed of electrically small ($\sim$ $\lambda_0$/2300) superconducting thin film resonators. The single resonant element is a planar spiral constructed by w= 3 $\mu$m wide semi-circular strips with increasing radii. The center to center separation between two adjacent turns is $s$= 6 $\mu$m. The spiral is densely packed with a large number of turns, N= 200, where the inner and outer diameters are D$_i$= 4.606 mm and D$_o$= 7 mm respectively. The electromagnetic response of the resonant spirals can be studied with a simple circuit model~\cite{BaenaSpiral}; i.e. an individual planar spiral can be treated as a magnetic resonator with an angular resonant frequency of $\omega_0$=1/$\sqrt{LC}$ where $L$ and $C$ are the total inductance and capacitance, respectively. 

Possessing large geometrical inductance and capacitance (due to large number of turns) over a compact geometry puts the individual spirals into a significantly sub-wavelength self-resonance regime. Therefore the metamaterial formed by the 3D array of these spirals can be considered as a continuous effective medium characterized with effective constitutive parameters; i.e. frequency dependent permeability $\mu_{\textrm{eff}}(\omega)$ and permittivity $\epsilon_{\textrm{eff}}(\omega)$. Previous theoretical~\cite{Gorkunov} and experimental~\cite{Massaoudi, Shadrivov} work on metamaterials with periodic circular conducting elements have shown that the fundamental resonant frequency of the metamaterial can be drastically lower than that of the single constitutive element mainly due to the mutual inductance (magnetic coupling) between the resonant elements. However, electrical (capacitive) coupling can also play a role in the collective electromagnetic response of the metamaterial, thus interaction between the individual elements can be complex depending on structural details of the metamaterial such as separation between the resonant elements along the substrate, interlayer spacing, array size and alignment/arrangement of the elements along the stacking direction~\cite{Balmaz, Powell}.  

Here we study a 3D superconducting metamaterial with sub-wavelength spiral elements operating in the rf regime. We find that the mutual interactions between the individual spirals contributes to the effective response of the macroscopic 3D metamaterial. Moreover, we observe that both the center frequency of the resonant bandwidth in the rf transmission and the negative $\mu_{\textrm{eff}}(\omega)$ region is highly temperature tunable.

\section{Experiment}

The metamaterial is patterned out of a 200 nm thick Nb thin film sputtered on to a 350 $\mu$m thick quartz substrate. We define circular spiral-shaped resonators~\cite{KurterAPL2010} into the Nb film by using standard photolithography and reactive ion etching using a mixture of CF$_4$/O$_2$ (10 $\%$ O$_2$). After fabrication, the quartz substrates are diced into 0.9~x~0.9 cm$^2$ and 2.6~x~5.2 cm$^2$ pieces to have individual spirals and 2D arrays of 3~x~6 Nb spirals. To be able to extend the structure into three dimensions, four substrates containing 2D arrays are stacked along the c-axis of the substrate by using thin 
sheets of Rohacell foam as spacers. The planes of the spirals are aligned in parallel and each substrate is separated from the next by 0.5 mm along the z-direction to produce strong coupling between the Nb spirals in the third dimension. The experiments have been conducted in an evacuated probe inside a cryogenic dewar and the temperature has been precisely adjusted by a temperature controller between 4.2 K up to well above the superconducting transition temperature, $T_c \simeq $ 9.2 K of Nb.

\section{Results and Discussion}

\begin{figure}
\centering
\includegraphics[bb=4 206 568 642,width=3.3 in]{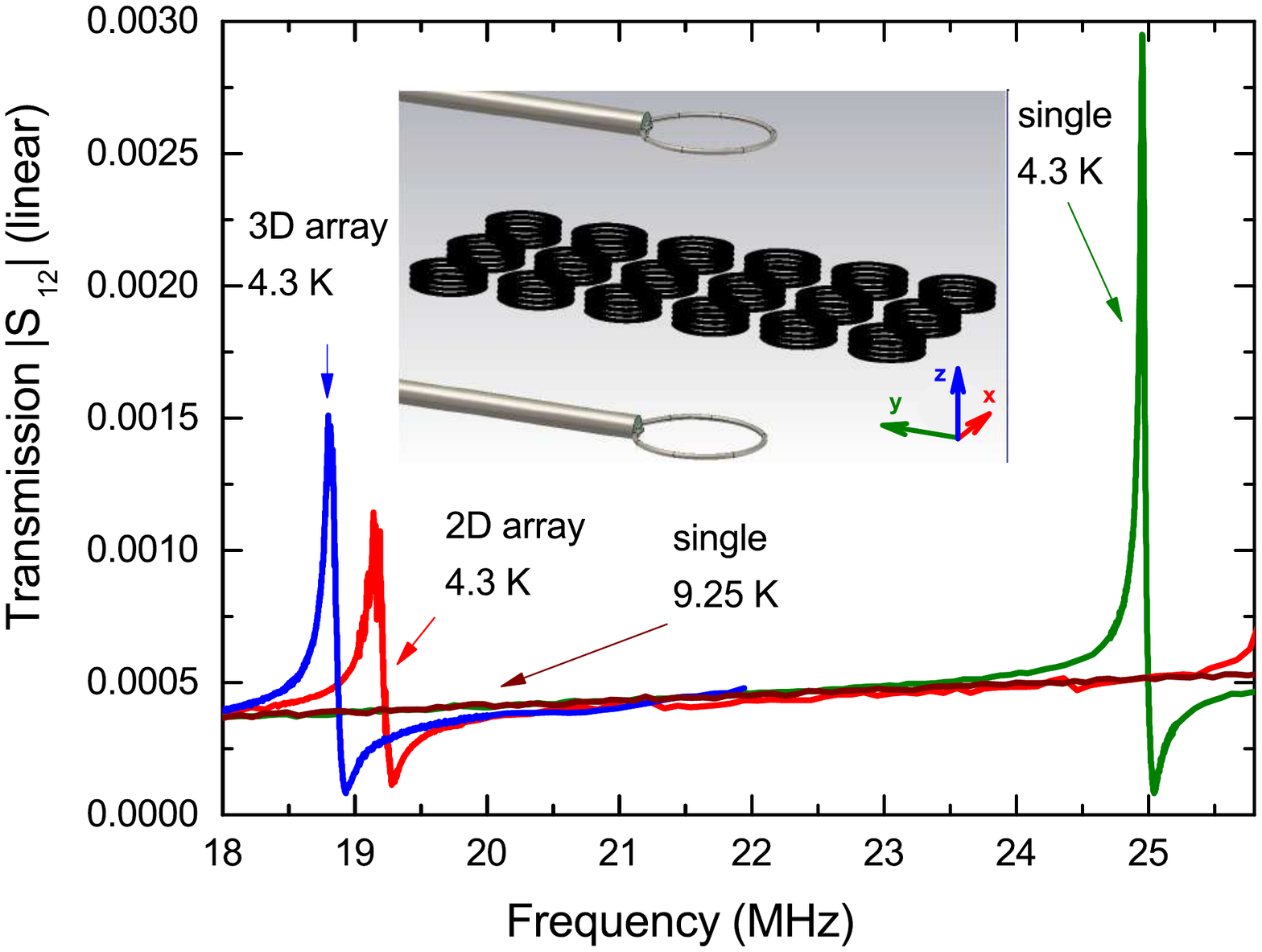}
\caption{(Color online) Comparison of transmission $|S_{12}|$ through a single spiral (green curve), 2D array of 3~x~6 spirals (red curve) and 3D array of the spirals formed by stacking four substrates including 2D arrays of spirals (blue curve). The lowest resonant frequency band decreases and develops fine splitting by adding more elements to the metamaterial. The maroon curve shows the data taken at the $T_c$ of Nb, 9.25 K, for a single spiral, where the resonant peak disappears due to enhanced ohmic losses. 2D and 3D arrays also show the same feature when the sample is heated above 9.25 K. The inset shows a schematic of the 3D metamaterial sandwiched between two rf-loop antennas which are 27.2 mm apart from each other along the z-axis.}
\label{comparison}
\end{figure}

Figure~\ref{comparison} shows the rf-transmission $|S_{12}|$ through a single Nb spiral along with the 2D and 3D arrays of the coupled spirals. The measurements have been performed by means of two loop antennas sandwiching the samples (see the sketch in the inset of Fig.~\ref{comparison} illustrating the 3D array of spirals between the antennas; the quartz substrates are not shown). The antennas are made of semi-rigid coax cables by shaping the inner conductors into 1.2 cm-diameter loops and attaching them to the outer conductors. The other ends of the coaxial cables are connected to a 2-port vector network analyzer to measure the scattering matrix. The magnetic excitation of the sample is achieved via inductive coupling into the top (drive) loop and the transmitted signal is picked up by the bottom (pickup) loop. 

The loop antennas are 27.2 mm apart to minimize the direct magnetic coupling between them~\cite{SpiralIEEE} that can obscure the response of the 3D metamaterial. The single spiral demonstrates a sharp fundamental resonance peak, $f_0$=$\omega_0$/2$\pi$ at $\sim$ 24.95 MHz and a dip at $\sim$ 25.04 MHz at 4.3 K with a loaded quality factor of 1273. The peak and the dip in the transmission spectrum correspond to constructive and destructive interferences between the direct coupling of the loops and magnetic coupling of the loops to the metamaterial respectively~\cite{BehnoodIEEE}. We have observed higher order modes corresponding to shorter-wavelength rf-current standing waves in the spiral, which have been extensively studied elsewhere~\cite{KurterAPL2010, PhysRevB.85.134535, BehnoodAPL, Ramaswamy201358, Maleeva2014}. Above the $T_c$ of Nb, superconductivity is lost and the resonance is wiped out at 9.25 K. This demonstrates that such a compact device is not functional if made with thin films of normal metals due to the large ohmic losses. An analytical model adapted to a similar experimental configuration showed that the transmission spectrum in the absence of the spiral resonator between the loops captures the experimental transmission through the spiral taken above the $T_c$ of the superconductor~\cite{BehnoodIEEE}.  

The low temperature resonance peak shifts to a significantly smaller frequency, 19.15 MHz in the rf-transmission spectrum of the 2D array of 3~x~6 spirals due to a significant change in effective parameters; especially $\mu_{\textrm{eff}}(\omega)$~\cite{Balmaz}. The small separation distance between the spirals $\sim$ $\lambda_0$/1875 strengthens the magnetic coupling between the individual spirals that leads to splitting of the resonant frequencies~\cite{RicciAPL2006}.

For 3D metamaterials with strong magnetic coupling in the stacking direction, the enhancement in capacitance due to interlayer interaction~\cite{PhysRevB.85.201104} further reduces the resonant frequency at 4.3 K down to 18.83 MHz as seen in the transmission spectrum of the 3D metamaterial.  The resonant features observed in the low temperature spectra of both 2D and 3D arrays of spirals disappear at temperatures above 9.25 K as expected. 

By considering the effective medium approximation, the frequency dependent complex effective permeability of an array of non-magnetic conducting SRRs has been characterized with a Lorentzian function as~\cite{Pendry99}

\begin{align}
\mu_{\textrm{eff}}(\omega)&= \mu'_{\textrm{eff}}(\omega)+i\mu''_{\textrm{eff}}(\omega) \nonumber \\
&=1-\frac {F\omega^2}{\omega^2-\omega_{r}^2+i\Gamma \omega}
\end{align}
where $\mu'_{\textrm{eff}}(\omega)$ and $\mu''_{\textrm{eff}}(\omega)$ are the real and imaginary parts of the effective permeability, $F$ is the filling fraction, $\omega_{r}$= 2$\pi f_r$ is the angular center frequency of the resonance, and $\Gamma$ is the damping term characterizing ohmic and radiation losses. The center frequency for the resonant bandwidth $f_r$ is determined by the resonant parameters of individual elements $f_0$, as well as the mutual interactions between them~\cite{Gorkunov, Lapine}.

The wavelength of the electromagnetic excitation we have used is much larger than the dimensions (D$_o$, w, $s$) and separation of the spirals forming our metamaterial. Therefore, we can apply the effective medium approximation and characterize the metamaterial with $\mu_{\textrm{eff}}(\omega)$ and $\epsilon_{\textrm{eff}}(\omega)$. Since inductive coupling is quite prominent here, the former is the dominant effective parameter. Sharp dips of the transmission spectra just above the resonant frequencies in Fig.~\ref{comparison} (below the background signal set by the normal state data taken at 9.25 K) imply the presence of negative permeability. To be able to characterize the negative magnetic response of our metamaterials, we have retrieved $\mu_{\textrm{eff}}(\omega)$ from the experimental scattering matrix~\cite{PhysRevB.85.201104} collected by two loops connected to a network analyzer (see the inset of Fig.~1). The induced electromotive force in the pickup loop as a function of frequency due to the magnetic coupling of top loop can be written as emf$(\omega)_{\textrm{ref}}$= -i$\omega$A$\mu_0H_z$ when there is no metamaterial between two loops and emf$(\omega)_{\textrm{meta}}$= -i$\omega$A$\mu_0\mu_{\textrm{eff}}(\omega) H_z$ when there is metamaterial sandwiched between them. Here, A is the area of the pickup coil, $\mu_0$ is the free space permeability and H$_z$ is perpendicular component of the magnetic field. By combining these two equations one can approximate the complex effective permeability as a ratio, $\mu_{\textrm{eff}}(\omega)$=emf$(\omega)_{\textrm{meta}}$/emf$(\omega)_{\textrm{ref}}$ $\sim$ $S_{21}(\omega)_{\textrm{meta}}$/$S_{21}(\omega)_{\textrm{ref}}$.  Here it is assumed that $H_z$ is the same in both measurements~\cite{PhysRevB.85.201104}.

\begin{figure}
\centering
\includegraphics[bb= 29 125 535 689,width=3.4 in]{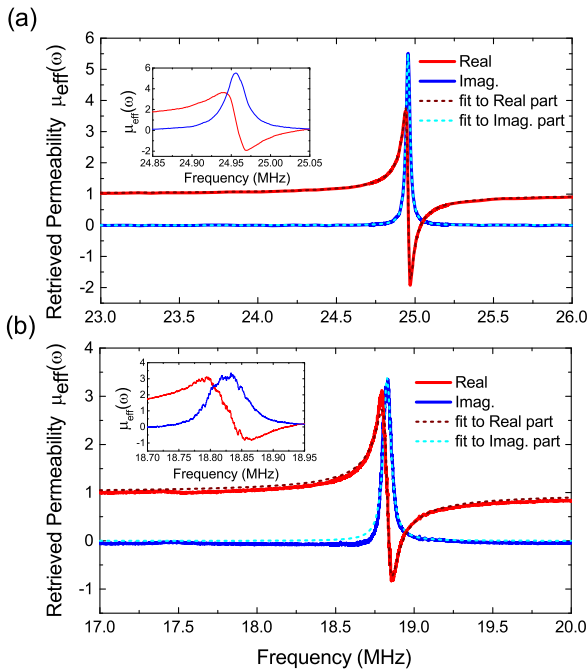}
\caption{(Color online) Real $\mu'_{\textrm{eff}}(\omega)$ and imaginary $\mu''_{\textrm{eff}}(\omega)$ parts of the effective complex permeability retrieved from the experimental scattering matrix at 4.3 K for (a) a single spiral, (b) 3D array of such spirals. The dashed maroon and cyan curves are Lorentzian fits to the retrieved data according to Eq.~(1) demonstrating a significant agreement. The insets are shorter frequency range plots of the data showing the details in the vicinity of the resonant frequency. A single smooth resonance peak of the individual spiral splits into multiple resonances for the 3D metamaterial.}
\label{retrieved}
\end{figure}

 Figure~\ref{retrieved} shows the real $\mu'_{\textrm{eff}}(\omega)$ (red curves) and imaginary $\mu''_{\textrm{eff}}(\omega)$(blue curves) parts of the retrieved permeability for (a) the single spiral and (b) the 3D metamaterial composed of similar spirals. The overlapping dashed curves are the Lorentzian fits according to Eq.~(1) which reasonably capture the retrieved data at 4.3 K. The fit parameters are $\omega_0 / 2\pi$= 24.955 MHz, $\Gamma/ 2\pi$= 0.03 MHz and $F$= 0.0068 for the single spiral and $\omega_r/ 2\pi$= 18.828 MHz, $\Gamma/ 2\pi$= 0.064 MHz and $F$= 0.012 for the 3D array. The enhanced loss in the 3D array compared to an individual spiral can be explained in part by the stronger inductive coupling to the excitation loops~\cite{Syms} and the excitation of magneto-inductive waves in the metamaterial~\cite{Tsironis}. We observe that the smooth well-defined resonance of the single spiral  splits into multiple resonances with magnetic coupling into other spirals in the 3D array (see the insets of the plots showing the resonant features in detail). Just above the resonant frequency, the $\mu'_{\textrm{eff}}(\omega)$ becomes negative in a narrow frequency band. These forbidden bands of frequencies range from 24.958 to 25.038 MHz for the single spiral and from 18.838 to 18.924 MHz for the 3D metamaterial at 4.3 K.

\begin{figure}
\centering
\includegraphics[bb=8 90 531 650,width=3.3 in]{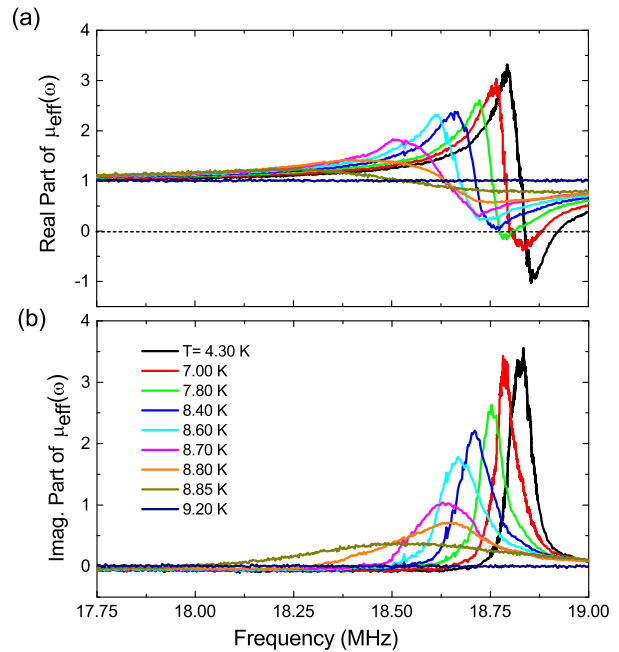}
\caption{(Color online) Temperature evolution of (a) real $\mu'_{\textrm{eff}}(\omega)$ and (b) imaginary $\mu''_{\textrm{eff}}(\omega)$ parts of the retrieved permeability for the 3D metamaterial. Both the frequency band where the $\mu'_{\textrm{eff}}(\omega)$ is negative and $f_r$ shift to smaller values of frequency with increasing temperature due to the enhanced inductance and losses.}
\label{temperature}
\end{figure}

Having discussed the low temperature magnetic response of the metamaterial, we now turn to the tunability of $\mu_{\textrm{eff}}(\omega)$ as the metamaterial is heated up to a temperature above the $T_c$ of Nb. Figure~\ref{temperature}(a) and (b) show the temperature dependence of the real and imaginary parts of the retrieved $\mu_{\textrm{eff}}(\omega)$ for the 3D metamaterial. The narrow frequency band where the $\mu'_{\textrm{eff}}(\omega)$ is negative shifts to lower frequency as the temperature increases as seen in Fig.~\ref{temperature}(a). The same effect is also seen in the peaks of the $\mu''_{\textrm{eff}}(\omega)$ that are often associated with the resonant frequencies~\cite{PhysRevLett.105.247402, PhysRevB.88.180510} (see Fig.~\ref{temperature}(b)). The frequency shift in the spectra is a result of a decrease in the density of superconducting electrons (superfluid density) with temperature $n_s(T)$ which is inversely proportional to the square of magnetic penetration depth $\lambda_L(T)$. As the fields of the electromagnetic waves penetrate into the metametarial deeper with increasing temperature, the total inductance of the individual spirals increases due to the enhanced geometrical and kinetic inductance, $L_k$ $\propto$ $\lambda_{L}(T)$ coth[t/$\lambda_{L}(T)$] where $t$ is the thickness of the Nb film used to make the metamaterial~\cite{Langley}. The enhanced total inductance leads to a reduction in the resonant frequency~\cite{Anlage1989}. Moreover, transmission decreases in magnitude and is smeared as $T_c$ is approached due to the enhanced ohmic losses arising from normal carriers, characterized by $\Gamma$ in Eq.~(1).

\begin{figure}
\centering
\includegraphics[bb=41 282 533 624,width=3.3 in]{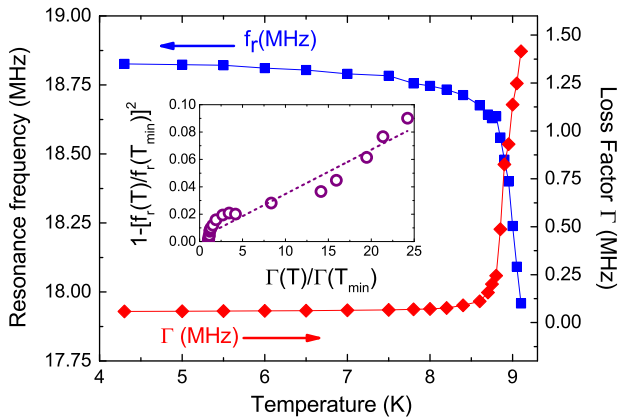}
\caption{(Color online) Temperature evolution of the extracted resonance frequency $f_r$ (blue squares) and damping factor $\Gamma$ (red diamonds) by fitting the retrieved permeability of 3D metamaterial according to Eq.~(1). In the vicinity of $T_c$, $f_r$ shows a steep drop as $\Gamma$ dramatically increases. The inset is a plot showing the quantities expected to be proportional to the normal fluid density $n_N(T)$.}
\label{extracted}
\end{figure}

The narrow frequency band in the real part of $\mu_{\textrm{eff}}(\omega)$ possesses negative values up to $\sim$ 8.3 K beyond which it starts to show near zero and then positive values. This crossover temperature corresponds to a similar value at which we start to see a discernible downward trend in the temperature evolution of $f_{r}(T)$ and an upturn in the damping term $\Gamma(T)$. To study this behavior, we extracted $\omega_r$ (thus $f_r$) and $\Gamma$ as a function of temperature through fitting the retrieved $\mu_{\textrm{eff}}(\omega)$ of the 3D metamaterial according to Eq.~(1). As seen in the main panel of Fig.~\ref{extracted}, for a long range of temperatures below $T_c$, both $f_r$ and $\Gamma$ are quite constant since there is no significant change in $n_s(T)$~\cite{Pambianchi}. As temperature increases, $n_s(T)$ starts to decrease leading to a weaker response in opposing the external magnetic field. Beyond 8.3 K, we see a slight decrease (increase) in $f_{r}$ ($\Gamma$) until the temperature is very near $T_c$. In the vicinity of $T_c$, we observe a steep drop (increase) in $f_r$ ($\Gamma$) mimicking the behavior of $n_s(T)$  ($n_{N}(T)$, normal fluid density) in this temperature regime. Here, $\lambda_{L}(T)$ and thus the total inductance dramatically increases (i.e. $L_k$ diverges) and accordingly $n_s(T)$ approaches zero. At $T_c$, ohmic losses are enhanced due to the dominance of $n_{N}(T)$ over $n_s(T)$ and resonant characteristics are completely wiped out. To demonstrate this more clearly and to test the validity of our extraction, we show 1-$[f_r(T)/f_r(T_{min})]^2$ vs. $\Gamma(T)$/$\Gamma(T_{min})$ in the inset of Fig.~\ref{extracted}. Both quantities are roughly proportional to the normal fluid density $n_{N}(T)$ so the plot should be approximately a straight line. Though we see a nonlinearity in the vicinity of the base temperature (due to the near absence of a frequency shift), the trend is quite linear at the higher temperatures, as expected.



\section{Summary}

We have developed a low-loss and ultra sub-wavelength 3D superconducting metamaterial. The engineered compact metamaterial shows a strong magnetic response to electromagnetic waves. The resonant frequency and frequency band with negative $\mu_{\textrm{eff}}(\omega)$ are precisely tuned by means of temperature without lumped elements. We observe a crossover from negative to positive $\mu'_{\textrm{eff}}(\omega)$ at a temperature near $T_c$ at which a dramatic increase occurs in losses ($\Gamma$) signaling a significant decrease in superfluid density $n_s(T)$. The work contributes to ongoing research on tunable negative permeability and near zero permeability metamaterials which can play an important role to improve the performance and directivity of ultra-small rf antennas.

\section{Acknowledgments}

This work was supported by the U.S. Office of
Naval Research through Grant No. N000140811058, the NSF-GOALI Program
through Grant No. ECCS-1158644, NSF-DMR 1410712, and the
Center for Nanophysics and Advanced Materials at the University
of Maryland. For the device fabrication, we acknowledge use of facilities of the Maryland Nanocenter at the University of Maryland. The authors would like to thank Brian Straughn and
John Abrahams for their techical assistance. We appreciate the helpful discussions with A.~D.~K. Finck, N. Maleeva, A. Yamilov.

\bibliography{3DmetamaterialREFS}

\end{document}